\documentclass[%
 reprint,
 amsmath,amssymb,
 aps,
 prl,
 longbibliography,
 lengthcheck,%
]{revtex4-1}

\usepackage{graphicx}% Include figure files
\usepackage{dcolumn}% Align table columns on decimal point
\usepackage{bm}% bold math
\usepackage{hyperref}% add hypertext capabilities
\begin{document}

%\preprint{APS/123-QED}

\title{Fano Mechanism of the Giant Magnetoresistance Formation in a Spin Nanostructure}% Force line breaks with \\

\author{Valery V. Val'kov$^{1,3}$}
 \email{vvv@iph.krasn.ru}
\author{Sergey V. Aksenov$^{1,2}$}%
 \email{asv86@iph.krasn.ru}
\affiliation{%
 $^{1}$L.V. Kirensky Institute of Physics, 660036 Krasnoyarsk, Russia\\
$^{2}$Siberian Federal University, 660041 Krasnoyarsk, Russia\\
$^{3}$Siberian State Aerospace University, 660014 Krasnoyarsk, Russia}%

\date{\today}% It is always \today, today,

\begin{abstract}
It is shown that, upon the electron quantum transport via the
nanostructure containing a spin dimer, the spin-flip processes
caused by the s-f exchange interaction between electron and dimer
spins lead to the Fano resonance effects. An applied magnetic
field eliminates degeneracy of the upper triplet states of the
dimer, changes the conditions for implementation of the Fano
resonances and antiresonances, and induces the new Fano resonance
and antiresonance. It results in the occurrence of a sharp peak
and dip in the energy dependence of transmittance. These effects
strongly modify the current--voltage characteristic of the
spin-dimer structure in a magnetic field and form giant
magnetoresistance.
\begin{description}
\item[PACS numbers]
73.23.-b, 73.63.Nm, 75.47.De, 75.76.+j, 71.70.Gm
\end{description}
\end{abstract}

\pacs{Valid PACS appear here}% PACS, the Physics and Astronomy
\maketitle

%\tableofcontents

In recent years, molecular compounds have been considered as
promising basis nanoelectronic elements playing an important role
in quantum transport~\cite{Fagas}. The exchange interaction
between spin moments of a conduction electron and a molecule
induces additional effects in the spin-polarized transport via
magnetic molecules. Among these effects are quantum tunneling of
magnetization of a molecule~\cite{Mis} and the Kondo
effect~\cite{Liang}. Such molecular compounds can exhibit strong
magnetic anisotropy sufficient to support stable orientation of
the spin at low temperatures~\cite{Gatt}. Therefore, along with
magnetic nanoheterostructures~\cite{Ved,Gul,Mir}, magnetic
molecular and atomic systems are currently considered as memory
elements. In view of this, the account for the
electron-electron~\cite{Del} and electron-phonon~\cite{Tikh,Ars}
interactions, which cause phase mismatch of conduction electrons
upon quantum transport, takes on great significance.

In addition to the mentioned features, it should be taken into
consideration that, upon coherent quantum transport via molecular
compounds, the transport properties of the latter can exhibit the
features related to the interference resonance between the states
of continuum and the states of a discrete spectrum (Fano
resonance)~\cite{Fano}. Currently, many resonance phenomena for a
great number of physical systems with discrete energy levels in
the continuum energy band are described well by the Fano formula
(see, for example, review~\cite{Kivshar}).

On this basis, it is reasonable to expect that spin degrees of
freedom and interactions between them offer new opportunities for
implementation of the Fano resonance phenomena in the coherent
spin-polarized transport via a spin nanostructure. One can suggest
that such resonance effects upon strong correlation between charge
and spin degrees of freedom manifest themselves, for example, in
giant magnetoresistance. The aim of this study was to confirm the
above statements by an example of the calculation of
current--voltage characteristics for a spacer containing
magnetoactive ions grouped in spin dimers.

\begin{figure}[htbp]
\begin{center}
\includegraphics[width=0.4\textwidth]{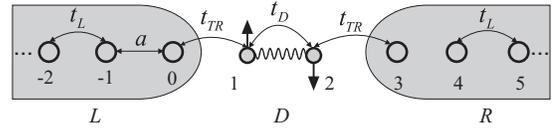}
\caption{Nanodevice in the form of the spin dimer
  connected with metal contacts.} \label{fig1}
\end{center}
\end{figure}

Consider a one-electron spin-polarized current via the region
containing a spin structure in the form of a spin dimer connected
with one-dimensional metal electrodes (Fig.\ref{fig1}). Such a
geometry of the problem corresponds to the case when the active
region represents an extremely thin layer of the material prepared
on the basis of a quasi-one-dimensional antiferromagnet, in which
chains of magnetic ions strongly coupled by the exchange
interaction are oriented perpendicular to the layer plane. This
situation is limit because a number of magnetic ions in a chain is
two, i. e., the dimer is formed.
The antiferromagnet is quasi-one-dimensional, which is revealed in
a negligibly small coupling of the spin dimers with one another.
Then, within the strong coupling method, hoppings in the
directions perpendicular to the dimer axis can be neglected, so
the electron transport becomes one-dimensional (Fig.\ref{fig1}).

Electron passage via the spin dimer structure exhibits the
features caused by the s-f exchange interaction between the spin
of a transferred electron and the spins of magnetoactive ions.
This is due to the fact that the s-f interaction not only induces
a potential profile of an electron flying throughout the structure
but also changes this profile by means of the spin-flip processes.
In this case, as will be shown below, resonance effects closely
related to the Fano effect can occur; their characteristics,
however, are directly related to the presence of spin degrees of
freedom in the system. Owing to this fact, it becomes possible to
attain giant values of magnetoresistance based on the Fano
resonance arising during the electron transport via a
quasi-molecular spin structure, which forms a potential profile
for the transferred electron depending on spin configuration of
the structure.

Below we will assume, as often, that the electrodes are connected
with reservoirs that are macroscopic conductors. Then, electrons
are thermalized and their temperature and chemical potential
coincide with those of the contact before their return to the
device, i. e., the contacts are considered to be reflectionless
and the electrodes, to be ideal~\cite{Bruus}. Over the entire
chain length, distance $a$ between sites is assumed to be
invariable.

We write the Hamiltonian of the system in the form
\begin{equation}\label{H}
\hat{H}=\hat{H}_{L}+\hat{H}_{R}+\hat{H}_{TR}+\hat{H}_{De}+\hat{H}_{sf}+\hat{H}_{D}+U_n,
\end{equation}
where two first terms correspond to the electron Hamiltonians of
the left ($L$) and  right ($R$) electrodes, which are expressed in
the secondary quantization representation as
\[
\hat{H}_{L}=\sum\limits_{\sigma, n=-\infty}^{0}
\biggl[\varepsilon_{L\sigma}c_{n\sigma}^{+}c_{n\sigma}+
t\left(c_{n\sigma}^{+}c_{n-1,\sigma}+c_{n-1,\sigma}^{+}c_{n\sigma}\right)\biggr],
\]
\[
\hat{H}_{R}=\sum\limits_{\sigma,
n=3}^{+\infty}\biggl[\varepsilon_{R\sigma}c_{n\sigma}^{+}c_{n\sigma}+
t\left(c_{n+1,\sigma}^{+}c_{n\sigma}+c_{n\sigma}^{+}c_{n+1,\sigma}\right)\biggr],
\]
where $c_{n\sigma}^{+}$ ($c_{n\sigma}$) is the creation
(annihilation) operator for a conduction electron with spin
$\sigma$ on site $n$ of electrode $\alpha$ ($\alpha=L,R$);
$\varepsilon_{\alpha\sigma}=\varepsilon_{\alpha}-g_e\mu_BH\sigma$
is the one-electron spin-dependent energy on the site of electrode
$\alpha$ in external magnetic field $H$, and $t$ is the hopping
integral (the same for both electrodes). The $Oz$ axis along which
the magnetic field is directed is perpendicular to the direction
of electron motion. The third term in expression \eqref{H}
describes hoppings of a conduction electron between the device and
the contacts:
\[
\hat{H}_{TR}=\sum\limits_{\sigma}t_{TR}\left(c_{1\sigma}^{+}c_{0\sigma}+c_{0\sigma}^{+}c_{1\sigma}+
c_{3\sigma}^{+}c_{2\sigma}+c_{2\sigma}^{+}c_{3\sigma}\right),
\]
and the fourth term of the Hamiltonian reflects the processes of
electron motion in the device:
\begin{eqnarray}\label{De}
&&\hat{H}_{De}=\sum\limits_{\sigma}\biggl[\varepsilon_{D\sigma}
\left(c_{1\sigma}^{+}c_{1\sigma}+c_{2\sigma}^{+}c_{2\sigma}\right)+\nonumber
\\
&&~~~~~~~~~~~~~~~~~~~~~~~~+t_{D}\left(c_{2\sigma}^{+}c_{1\sigma}
+c_{1\sigma}^{+}c_{2\sigma}\right)\biggr].\nonumber
\end{eqnarray}

The interaction between the transferred electron and the dimer
spins is determined by the term
\begin{eqnarray}\label{Hsf}
&&\hat{H}_{sf}=\frac{A_{sf}}{2}\sum\limits_{n=1}^{2}
\biggl[\left(c_{n\uparrow}^{+}c_{n\downarrow}\hat{S}_{n}^{-}+
c_{n\downarrow}^{+}c_{n\uparrow}\hat{S}_{n}^{+}\right)+\nonumber
\\
&&~~~~~~~~~~~~~~~~~~~~~~~~+\left(c_{n\uparrow}^{+}c_{n\uparrow}-
c_{n\downarrow}^{+}c_{n\downarrow}\right)\hat{S}_{n}^{z}\biggr].
\end{eqnarray}
Here, $A_{sf}$ is the parameter of the s-f exchange interaction
and $\hat{S}_{n}^{+},~\hat{S}_{n}^{-}$ and $\hat{S}_{n}^{z}$ are
the operators of the spin moment entering the dimer structure and
located on site $n$.

Operator $\hat{H}_{D}$ in expression~\eqref{H} describes the
exchange interaction between the dimer spin moments and their
energy in magnetic field $H$:
\begin{equation}\label{HDd}
\hat{H}_{D}=I\left(\mathbf{S_{1}}\mathbf{S_{2}}\right)-g
\mu_{B}H\left(S_{1}^{z}+S_{2}^{z}\right),
\end{equation}
where $I>0$ is the parameter of the exchange interaction. Below,
we will consider the case of a weak magnetic field
($g\mu_{B}H<I$); therefore, the singlet state will be the ground
state of the dimer. The last term in the Hamiltonian characterizes
the potential energy of an electron in an electric field. It is
assumed that there is the potential jump $V=V(2)-V(1)>0$ between
the dimer sites such that $V(1)=V(-\infty < n \leq 0)$ and
$V(2)=V(3 \leq n < \infty)$.

In solving the Schrodinger equation, due to the presence of the
s-f exchange interaction, one must take into account the states
with different spin projections for both the transferred electron
an the dimer. As basis states, we choose the states characterizing
the spin state of the transferred electron on site $n$ and one of
the four states of the spin dimer:
$D_{JJ_z}c_{n\sigma}^{+}|0\rangle$; $D_{00}$ describes the singlet
state of the dimer and $~D_{11}$, $~D_{10}$ and $~D_{1,-1}$
describe its triplet states $(J=1)$ with $~J_z=1$, $~J_z=0$ and
$~J_z=-1$, respectively. Here~$|0\rangle$ is the vacuum state for
the fermion subsystem.

Let us consider the case when the electron approaching the device
from the left contact has the spin projection $\sigma^{z}=1/2$ and
the spin dimer is in the singlet state $D_{00}$. Then, the wave
function of the system is written as
\begin{equation}\label{PsiL}
|\Psi\rangle=\sum_n\biggl[w_{n}c_{n\uparrow}^{+}D_{00}+
u_{n}c_{n\uparrow}^{+}D_{10}+
v_{n}c_{n\downarrow}^{+}D_{11}\biggr]|0\rangle.
\end{equation}
For the electron injected by the left contact with wave vector
$k_{L}$, the expansion coefficients in the left $(n\leq0)$ and
right $(n\geq3)$ contacts has the form
\begin{eqnarray} \label{wuvL}
&&n\leq0:~w_{n}=e^{ik_{L}n}+r_{00}e^{-ik_{L}n};\nonumber
\\
&&~~~~~~~~~~~u_{n}=r_{10}e^{-iq_{L}n};~v_{n}=r_{11}e^{-ip_{L}n};
\\
&&n\geq3:~w_{n}=t_{00}e^{ik_{R}n};\nonumber
\\
&&~~~~~~~~~~~u_{n}=t_{10}e^{iq_{R}n};~v_{n}=t_{11}e^{ip_{R}n},\nonumber
\end{eqnarray}
where $r_{00}~ (t_{00}),~r_{10} (~t_{10})$ and $r_{11}~(t_{11})$
are the reflection (transmission) amplitudes when the dimer is in
the singlet and triplet states, respectively, and $k_{L},~k_{R}$,
$q_{L},~q_{R}$ and $p_{L},~p_{R}$ are the wave vectors satisfying
the relations
\begin{eqnarray}\label{BandsLR}
&&E=U_{1(2)}+2|t|(1-cosk_{L(R)});\nonumber
\\
&&E=U_{1(2)}+I+2|t|(1-cosq_{L(R)});
\\
&&E=U_{1(2)}+I+(2-g)\mu_BH+2|t|(1-cosp_{L(R)}).\nonumber
\end{eqnarray}
These relations are written with the change in electron energy
count by the value $-2|t|-3I/4-\mu_{B}H$ and on the assumption
that the left and right conductors are identical. In the absence
of bias voltage, $k_L=k_R=k$, $q_L=q_R=q$, and $p_L=p_R=p$.

From the Schrodinger equation, we obtain the set of 12 equations
for six reflection and transmission amplitudes and six
coefficients $w_1,  w_2,  u_1,  u_2,  v_1$, and $v_2$. After
certain transformations, this system is reduced to the three
equations containing only the transmission amplitudes and, at
$V=0$, has the form
\begin{eqnarray}\label{MatrSys}
&&\left[1+3A_{sf}^2-\left(e^{-ik}+4\varepsilon_{D}\right)^2\right]e^{2ik}t_{00}+A_{sf}B_{1}e^{2iq}t_{10}-\nonumber
\\
&&~~~~~~~~~~~~~~~~~~~~~~~~~~~~~~~-\sqrt{2}A_{sf}B_{2}e^{2ip}t_{11}=e^{2ik}-1,\nonumber
\\
&&-A_{sf}B_{1}e^{2ik}t_{00}+\left[1-A_{sf}^2-\left(e^{-iq}+4\varepsilon_{D}\right)^2\right]e^{2iq}t_{10}-\nonumber
\\
&&~~~~~~~~~~~~~~~~~~~~~~~~~~~~~~~-\sqrt{2}A_{sf}B_{3}e^{2ip}t_{11}=0,
\\
&&\sqrt{2}A_{sf}B_{2}e^{2ik}t_{00}-\sqrt{2}A_{sf}B_{3}e^{2iq}t_{10}+\nonumber
\\
&&~~~~~~~~~~~~~~~~~~~+\left[1-\left(e^{-ip}-A_{sf}+4\varepsilon_{D}\right)^2\right]e^{2ip}t_{11}=0,\nonumber
\end{eqnarray}
where $B_{1}=e^{-ik}-e^{-iq}+2A_{sf}$,
$B_{2}=e^{-ik}-e^{-ip}+2A_{sf}$,
$B_{3}=e^{-iq}+e^{-ip}+8\varepsilon_{D}$. Hereinafter, all the
energy quantities will be measured in the bandwidth units
$W=4\left|t\right|$.

We limit the consideration to the low-energy region ($E < I$),
where the electron transport is implemented along one channel,
since the kinetic energy of an electron is insufficient to
transfer the dimer to the triplet states. This energy range is of
interest because within this range, as will be shown below, the
Fano resonance effects occur that allow implementation of giant
magnetoresistance.

\begin{figure}[htbp]
\begin{center}
\includegraphics[width=0.4\textwidth]{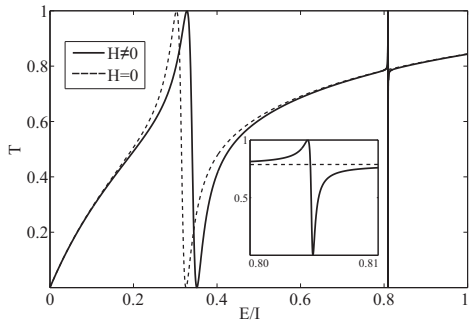}
\caption{Modification of the energy dependence of the electron
transmittance at the switched magnetic field. The insert shows the
structure of the induced asymmetric peak. $t=-0.2$ eV,
$I=A_{sf}=0.125$, $\mu_{B}H=0.005$, $\varepsilon_{D}=-0.125$,
$g=1$, $V=0$.} \label{fig2}
\end{center}
\end{figure}

At $E < I$, transmittance $T$ is determined only by amplitude
$t_{00}$: $T=\mid t_{00}\mid^2$. Therefore, solving set
\eqref{MatrSys}, we arrive at
\begin{equation}\label{TGeneral}
T=\frac{4\gamma^2\sin^2k}{[\gamma b_2-2A_{sf}^2a_1b_1]^2+[\gamma
a_2+A_{sf}^2(b_1^2-a_1^2)]^2},
\end{equation}
where
\begin{eqnarray}\label{const}
&&\gamma=A_{sf}\delta_p (\omega_q+\omega_p)^2+A_{sf}^2+
\omega_q^2-1,\nonumber\\
&&\omega_{q}=4\varepsilon_{D}+e^{-iq},~\delta_p=2A_{sf}\left(1-\left(\omega_p-
A_{sf}\right)^2\right)^{-1},\nonumber\\
&&\omega_{p}=4\varepsilon_{D}+e^{-ip},~a_1=\delta_p
(\omega_q+\omega_p)\lambda_p(k)-\lambda_q(k),\nonumber\\
&&\lambda_{q}\left(k\right)=2A_{sf}+\cos k - e^{-iq},\nonumber\\
&&b_1=\left[1-\delta_p (\omega_q+\omega_p)\right]\sin k,~~\\
&&a_2=1+3A_{sf}^2-\left(4\varepsilon_{D}+cosk\right)^2+\left(1-A_{sf}\delta
\right)\sin^2k+\nonumber\\
&&+A_{sf}\delta_p
\lambda^2_{p}\left(k\right),\nonumber\\
&&b_2=2\sin k \left[4\varepsilon_{D}+\cos k -A_{sf}\delta_p
\lambda_{p}\left(k\right)\right].\nonumber
\end{eqnarray}
At zero magnetic field, $H=0$, the transmittance is written in the
simpler form
\begin{equation}
\label{T0}
T=\frac{4G_{q}^2\sin^2k}{4Z_{q}^2\left(k\right)\sin^2k+R_{q}^2\left(k\right)}.
\end{equation}
Here, we use the following notation:
\begin{eqnarray}\label{GZRl}
&&G_{q}=\left(\omega_{q}-2A_{sf}\right)^2-3A_{sf}^2-1,\nonumber\\
&&Z_{q}\left(k\right)=G_{q}\left[\omega_{q}-2A_{sf}+
\lambda_{q}\left(k\right)\right]+3A_{sf}^2\lambda_{q}\left(k\right),\nonumber\\
&&R_{q}\left(k\right)=\left(G_{q}+3A_{sf}^2\right)\left[\sin^2k-\lambda_{q}^2\left(k\right)\right]+\nonumber\\
&&+G_{q}\left[1+3A_{sf}^2+\lambda_{q}^2\left(k\right)-
\left(\omega_{q}-2A_{sf}+\lambda_{q}\left(k\right)\right)^2\right],\nonumber
\end{eqnarray}
The energy dependence of transmittance at $H=0$ is shown in
Fig.\ref{fig2} by the dotted line. It can be seen that the
monotonic growth of $T$ with increasing kinetic energy of electron
from zero value is changed for the splash of $T$ up to the maximum
value and the sharp drop down to the zero value. After that, the
$T(E)$ dependence monotonically increases again. Total
transmission $(T=1)$ and total reflection $(T=0)$ correspond to
the Fano resonance and antiresonance~\cite{Fano} and are related
to interference of the continuous spectrum states and discrete
spectrum states. The energy value at which the antiresonance
occurs is found from the condition of vanishing $G_{q}$:
\begin{equation}
\label{Emin}
E=I-\frac{\left(\alpha-1\right)^2}{4\alpha},~\alpha=2A_{sf}-4\varepsilon_{D}+\sqrt{3A_{sf}^2+1},
\end{equation}
Near the same value, the Fano resonance is located at $T=1$.

In a magnetic field, the triplet states of the spin dimer are
split, which shifts the values of the discrete spectrum energies.
Therefore, apart from the shift of the above-mentioned Fano
resonance and antiresonance points (solid curve in
Fig.\ref{fig2}), at $H \ne 0$ the new, magnetic-field-induced Fano
resonance and antiresonance can occur. In Figure \ref{fig2}, this
effect is revealed in the very narrow peak and dip of the $T(E)$
dependence in the vicinity of $E/I \simeq 0.8 $. The
magnetic-field-induced Fano resonance and antiresonance is
enlarged in the insert in Fig. \ref{fig2}.

Mathematically, the above statements follow from the analysis of
the expression for $T(E, H)$ at $H\longrightarrow0$. For the $H
\ne 0$, the transmittance can be expressed as
\begin{equation}
\label{TH} T=\frac{4\left[G_{q}\Omega_{q}+\Delta \cdot
\Phi\right]^2 \sin^2k} {4\left[\Omega_{q}Z_{q} + \Delta \cdot
F\right]^2 \sin^2k+\left[\Omega_{q}R_{q}+ \Delta
\cdot\Psi\right]^2},
\end{equation}
where
\[
\Delta=\omega_{p}-\omega_{q},~~\Omega_{q}=1-\left(\omega_{q}+
A_{sf}\right)^2.
\]
The specific form of $\Phi, F$, and $\Psi$ can be easily obtained
from comparison of expressions \eqref{TGeneral} and \eqref{TH};
however, in our case, it is not important. It is significant only
that the condition imposed on the magnetic-field-induced Fano
antiresonance acquires the form
\begin{equation}\label{RelatH}
G_q\Omega_q=-\Delta\Phi.
\end{equation}
If $\Delta=0$, this condition is met at $G_q=0$ or at
$\Omega_q=0$. However, the second equation does not lead to the
antiresonance, since there are the same factors in the denominator
of \eqref{TH}.

If $\Delta \ne 0$, then, due to the difference in the expressions
for $\Phi, F$, and $\Psi$, there are no reducible factors and the
second solution can be implemented. This corresponds to inducing
the Fano resonance and antiresonance by a magnetic field.

To sum up at the end of this section, note that the formation of
the Fano resonance effects in this system is due to the spin-flip
processes, since, as the calculation showed, there are no
resonance effects when the exchange couplings are described by the
Ising (not Heisenberg) Hamiltonians.

In order to calculate the current--voltage characteristic, we use
the Landauer--Buttiker method~\cite{Bruus,Datta}. Within this
method, the dependence of current on applied voltage is determined
by the well-known expression
\begin{equation}\label{I}
I(V)=\frac{e}{h}\int dE \bigl[T(E) f_{L}(E)-T'(E) f_{R}(E)\bigr],
\end{equation}
where $f_{L}(E) \equiv f(E-\mu_L)$ and $f_{R}(E) \equiv
f(E-\mu_R)$ are the Fermi functions of the electron distribution
in the left and right contacts with the electrochemical potentials
$\mu_L=E_F$ and $\mu_R=E_F-eV$, respectively; $T(E)$ is the
above-mentioned transmittance for the electron injected from the
left electrode; and $T'(E)$ is the transmittance for the electron
transferred from the right electrode. The coefficients $T'(E)$ and
$T(E)$ are calculated similarly.

\begin{figure}[htbp]
\begin{center}
\includegraphics[width=0.4\textwidth]{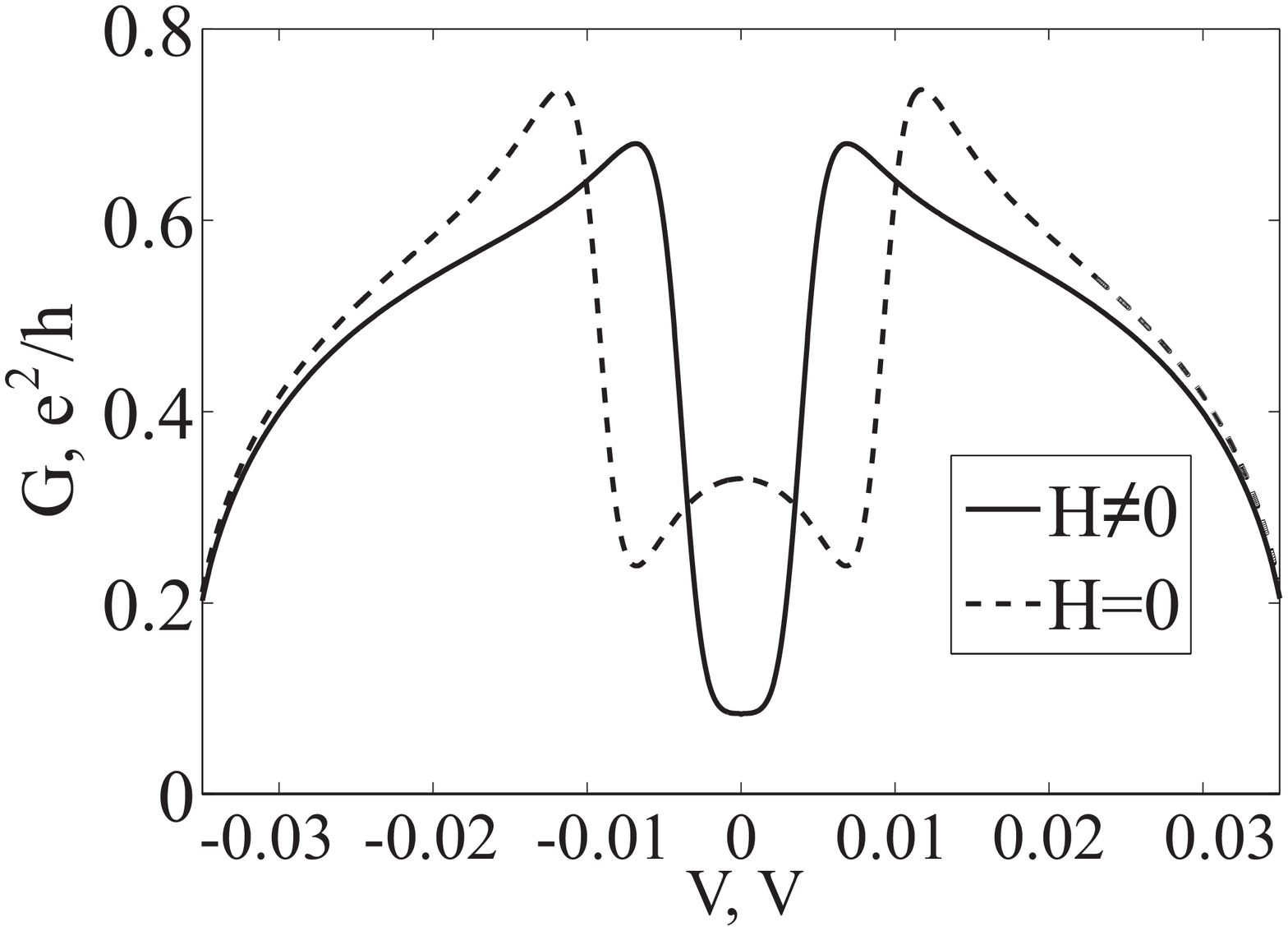}
\caption{Differential conductance of the spin-dimer structure. The
parameters of the system are the same as in Fig.\ref{fig2},
$E_{F}\simeq0.045$, $T\simeq0.3$ K.} \label{fig3}
\end{center}
\end{figure}

Figure \ref{fig3} demonstrates the variation in the dependence of
the differential conductance $G=\partial I(V)/\partial V$ measured
in the conduction quantum ($e^2/h$) units on bias voltage $V$ at
the switched magnetic field. The parameters of the system are such
that at $H=0$ the electron energies giving the main contribution
to conductance are localized near the antiresonance on the right.
With increasing $V$, the antiresonance becomes more pronounced and
conductance drops. With a further increase in $V$, the Fano
resonance starts contributing; correspondingly, the differential
conductance (dotted line) becomes nonmonotonic.
\begin{figure}[htbp]
\begin{center}
\includegraphics[width=0.4\textwidth]{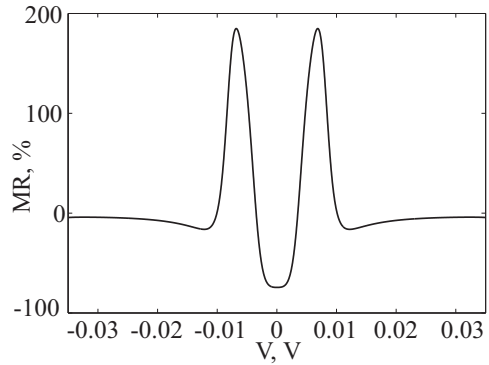}
\caption{Magnetoresistance at the same parameters as in Fig.
\ref{fig3}.} \label{fig4}
\end{center}
\end{figure}
Switching off the magnetic field shifts the Fano resonance and
antiresonance (Fig. \ref{fig2}). The magnetic field may appear
such that the transmittance at $V=0$ is minimum (the Fano
antiresonance case). It results in such a modification of the
dependence of the differential conductance on bias voltage that at
the switched magnetic field, in the region of small $V$, the
maximum $G$ changes for the minimum $G$ (solid curve in Fig.
\ref{fig3}). These effects cause the occurrence of the large
magnetoresistance $MR=(G(H)/G(0)-1)\times100\%$ at the switched
magnetic field. The bias dependence of magnetoresistance is
presented in Fig. \ref{fig4}. This result clearly demonstrates the
possibility of attaining large magnetoresistance values by
shifting the Fano resonance and antiresonance in the magnetic
field. Magnetoresistance can take negative values.
\begin{figure}[htbp]
\begin{center}
\includegraphics[width=0.4\textwidth]{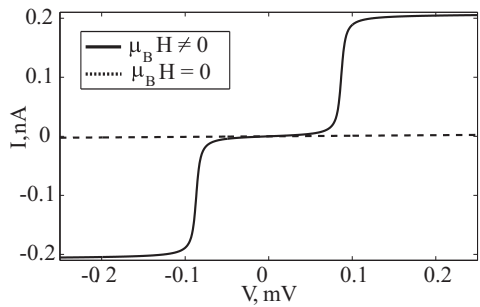}
\caption{Effect of the magnetic field on the current--voltage
characteristic. The parameters of the system are
$t_{L}=t_{R}=-0.1$~eV, $t_{TR}=-0.125$, $t_{D}\simeq-0.186$,
$\mu_{B}H=6.25\times10^{-3}$, $I=0.187$, $A_{sf}=0.75$,
$\varepsilon_{D}=-0.525$, $g=1$, $E_F\simeq 0.13$ and $T\simeq3$
mK.} \label{fig5}
\end{center}
\end{figure}

\begin{figure}[htbp]
\begin{center}
\includegraphics[width=0.4\textwidth]{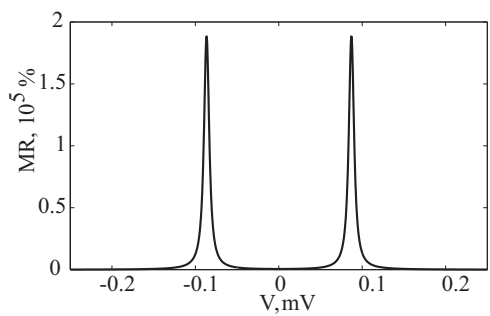}
\caption{Magnetoresistance at the same parameters of the system as
in Fig. \ref{fig5}.} \label{fig6}
\end{center}
\end{figure}

The magnetic-field-induced Fano resonance can cause abnormally
large magnetoresistance values. Of the most interest is the case
when the characteristic electron energies determining the
resistive characteristics were localized in the vicinity of the
Fano resonance in zero magnetic field. The parameters of the
system and the external magnetic field can be such that at $H \neq
0$ the Fano resonance occurs. Then, one should expect the strong
conductance growth. Figures \ref{fig5} and \ref{fig6} demonstrate
the current--voltage characteristic and magnetoresistance for this
case. The current jump at $|V|\simeq 0.1$ mV (solid line in Fig.
\ref{fig5}) arises because the resonance energy state starts
contributing to the current-carrying states. It can be seen that
the resistance variation in the magnetic field can reach $10^5\%$
(Fig. \ref{fig6}).

In this study, as an active element located between the metal
electrodes, we used a nanostructure containing a spin dimer. The
s-f exchange interaction between the conduction electron spin and
the dimer spins forms a potential profile that contains both the
constant (the Ising part of the interaction) and fluctuating
(transversal part of the Heisenberg form $\sim S^+\sigma^-
+S^-\sigma^+$) components. Owing to the fluctuating part of the
interaction, the Fano resonance effects occur.

For practical application, two effects caused by the magnetic
field are important. The first effect is related to the shift in
the energy region of the Fano resonance and antiresonance. The
second, brighter effect results from splitting of the upper
high-spin states. Due to the occurrence of the modified energy
values of the discrete states in the system, these split states of
the discrete spectrum also exhibit interference. It results in
induction of the new Fano resonance and antiresonance. These
effects all together cause the large values of magnetoresistance.
Since the above statements on the effect of magnetic field are
general, one can expect that the analogous effects of the giant
magnetoresistance formation take place in other spin
nanostructures.

This study was supported by the Program of the Physical Sciences
Division of the Russian Academy of Sciences; Federal target
program Scientific and Pedagogical Personnel for Innovative
Russia, 2009-2013; Siberian Branch of the Russian Academy of
Sciences, interdisciplinary project no. 53; Russian Foundation for
Basic Research, project no. 09-02-00127, no. 11-02-98007. One of
authors (S.A.) would like to acknowledge the support of the grant
MK-1300.2011.2 of the President of the Russian Federation.

\end{document}